\documentclass[aps,pre,amsmath,amssymb,showpacs]{revtex4}
\pdfoutput=1
\usepackage{amsmath}
\usepackage{amssymb}%
\usepackage{amsfonts}%
\usepackage{graphicx}

\begin{document}

\title{Beyond the Heisenberg time: Semiclassical treatment of spectral correlations in chaotic systems with
spin 1/2 }
\author{ Petr Braun}
\affiliation{{  Fachbereich Physik, Universit\"at Duisburg--Essen,
47048 Duisburg,
  GERMANY\\
   Institute of Physics, Saint-Petersburg University,
198504 Saint-Petersburg,  RUSSIA}}

\begin{abstract}
The two-point  correlation function of chaotic systems with spin
1/2 is evaluated using  periodic orbits. The spectral form factor
for all times thus becomes accessible. Equivalence with the
predictions of  random matrix theory for the Gaussian symplectic
ensemble is demonstrated. A duality between the underlying
generating functions of the orthogonal and symplectic symmetry
classes is semiclassically established.

\end{abstract}

\maketitle
\section{\bigskip INTRODUCTION}

Energy levels of classically chaotic systems exhibit correlation
only slowly subsiding with the energy offset \cite{Haake10}.
Around 1980 it became clear, after extensive numerical
experiments, that with as few as two degrees of freedom, spectral
correlations in the highly excited energy domain have universal
properties and obey the same laws as the eigenvalues of the
Gaussian random matrix ensembles of the appropriate symmetry class
\cite{Bohig84, McDon79, Casat80, Berry87}. This assertion, known
as the Bohigas-Giannoni-Schmitt (BGS) conjecture took a
surprisingly long time to be proven. In most cases the tool used
was the Gutzwiller formula giving the spectral density $\rho\left(
E\right)  $ in a chaotic system as sum over the classical
periodic orbits, each orbit creating a contribution $\sim e^{iS_{\gamma}%
/\hbar}$ where $S_{\gamma}$ is the action of the orbit $\gamma$.
Its substitution into the spectral correlation function and its
Fourier transform $K\left( \tau\right)  $,  the spectral form
factor, leads to double sums over orbit pairs with summands
proportional to $e^{i\left(  S_{\gamma}-S_{\gamma^{\prime
}}\right)  /\hbar}$. Significant contributions can be expected
only from pairs with the action difference not large compared with
$\hbar$.

The first success in the proof of BGS was connected with the
diagonal approximation \cite{Berry85} which takes into account
only pairs with $S_{\gamma}=S_{\gamma'}$; it explained the fact
that at small times $K\left(  \tau\right)  \approx2\tau$ (time
reversal allowed, orthogonal universality class) or $\tau$ (time
reversal forbidden,  unitary class). Fifteen years later came the
realization that a long periodic orbit with a small-angle
self-crossing dividing it into two pieces, has a `` partner''\
orbit with the crossing avoided, but otherwise almost unchanged,
up to the sense of traversal of one of the pieces \cite{Siebe01}.
Contributions of such `` Sieber-Richter pairs''\ sum up the
next-to-leading term $-2\tau^{2}$ in the form factor. Summation
over pairs in which the partner consists of pieces of the original
orbit reconnected in all thinkable ways, restores the small-time
form factor to all orders \cite{Muell04,Muell05}.

The form factor experiences a break-up of analyticity at $\tau=1$,
i.e., at the Heisenberg time $T_{H}=2\pi\hbar\bar{\rho}$, which
reflects the existence of an oscillatory component of the
correlation function with the period of the mean level spacing
$\Delta=1/\bar{\rho}$. That component is overlooked in the
straightforward semiclassical approach providing correlation
functions as asymptotic power series in $1/\varepsilon$; the
reason is that semiclassical sums need for their convergence a
non-vanishing positive imaginary part of the energy parameters,
however
$e^{i\varepsilon}\sim0\cdot\varepsilon^{-1}+0\cdot\varepsilon
^{-2}+\ldots$ if $\operatorname{Im}\varepsilon>\delta>0$. Early
estimates of the oscillatory components are contained in
\cite{Bogom96a,Keppe00}. Systematic approach is based on the
formalism of generating functions, i.e., averaged ratios of the
spectral determinants. In this approach partnership of more than
two classical orbits is taken into account, and use is made of the
semiclassical approximation of the spectral determinant known as
the Riemann-Siegel look-alike \cite{Berry90,
Keati92,Berry92,Keati07}. As a result complete agreement of the
semiclassical correlation functions with RMT for spinless systems
was demonstrated \cite{Heusl07,Muell09}.

Systems with half-integer spin belong to the symplectic
universality class whose RMT counterpart is the Gaussian
symplectic ensemble. The spin coupling to  chaotic translational
motion leads to randomization of the spin evolution
\cite{Keppe10}. The  ergodicity of that  evolution is instrumental
for the evaluation of the relevant periodic orbit expansions such
as the diagonal sum for the form factor \cite{Bolte99} and the
contribution of the Sieber-Richter pairs which changes its sign in
the presence of a half-integer spin \cite{Heusl01}. The full
expansion of the form factor of systems with symplectic symmetry
for times smaller than $T_{H}$ was obtained for the quantum graphs
in \cite{Bolte03} and for general dynamical systems in
\cite{Muell05}.

Here we close the gap in the proof of BGS for the systems with
half-integer spin by showing the equivalence of their
semiclassical correlation function with the RMT predictions
including the oscillatory terms; the corresponding form factors
coincide for all times. Analytical properties of the correlation
function are used to recover the oscillatory term with smaller
frequency responsible for the well-known logarithmic singularity
of the form factor. The derivation employs the
 simple duality discovered between
the semiclassical 4-determinant generating functions of chaotic
systems with  symplectic and orthogonal symmetry. We talk mostly
about the spin $1/2$ case although the results remain true for
other half-integer spins.

\section{Complex correlator, form factor and generating function}

Our main object of study will be the complex two-point spectral
correlation function (complex correlator, for short). This is an
analytic function of the complex dimensionless variable
$\varepsilon$ which is the double spectral sum,
\begin{equation}
C\left(  \varepsilon\right)  =\frac{\Delta^{2}}{2\pi^{2}}\left\langle
\sum_{i\neq k}\frac{1}{\left(  E_{k}-E-\frac{\varepsilon\Delta}{2\pi}\right)
}\frac{1}{\left(  E_{i}-E+\frac{\varepsilon\Delta}{2\pi}\right)
}\right\rangle -\frac{1}{2}\label{defcorr}%
\end{equation}
where $\Delta$ stands for the mean level spacing;
$\langle\ldots\rangle$ denotes averaging over an interval of the
reference energy $E$,  classically small but large compared with
$\Delta$. The complex correlator is defined in the half plane
$\operatorname{Im}\varepsilon>0$ where it is analytic and tends to
zero when $|\varepsilon|\rightarrow\infty$, and can be continued
to the lower half plane where (\ref{defcorr}) would no longer be
true. The real part of $C\left(  \varepsilon\right)  $ at the
positive real axis coincides with the real level-level correlation
function \cite{Haake10} while its Fourier transform is the
spectral form factor $K(\tau)$; the connection between the two
functions is given by
\begin{eqnarray}
K\left(  \tau\right)   &
={\cal{F}}(C)=\frac{1}{2\pi\tau_{H}}\int_{-\infty+i0}^{\infty
+i0}C\left(  \varepsilon\right)  e^{-i2\varepsilon\tau/\tau_{H}}%
d\varepsilon,\quad\tau>0;\label{CtoKK}\\%
C\left(  \varepsilon\right)   &
={\cal{F}}^{-1}(K)=2\int_{0}^{\infty}e^{i2\tau\varepsilon
/\tau_{H}}K\left( \tau\right)
d\tau,\quad\operatorname{Im}\varepsilon
>0.\label{KtoC}
\end{eqnarray}
Here $\tau_{H}$ stands for $1$\ for the orthogonal universality
class and $2$ for the symplectic class; such choice is equivalent
to the replacement $2\varepsilon\rightarrow\varepsilon$ in the
symplectic case motivated by the Kramers degeneracy
\cite{Keppe00}.

The semiclassical evaluation of the complex correlator is based on
the
 generating function defined as the averaged ratio of four
spectral determinants,
\begin{equation}
Z\left(  \hat{\varepsilon}\right)  =\left\langle \frac{\det\left(
H-E-\frac{\varepsilon_{C}\Delta}{2\pi}\right)  \det\left(  H-E-\frac
{\varepsilon_{D}\Delta}{2\pi}\right)  }{\det\left(  H-E-\frac{\varepsilon
_{A}\Delta}{2\pi}\right)  \det\left(  H-E-\frac{\varepsilon_{B}\Delta}{2\pi
}\right)  }\right\rangle .\label{defZ}%
\end{equation}
where $\left(  \hat{\varepsilon}\right)  \equiv\left(  \varepsilon
_{A},\varepsilon_{B,}\varepsilon_{C},\varepsilon_{D}\right)  $.The complex
correlator can be obtained from $Z$ as
\begin{equation}
C\left(  \varepsilon\right)  =\lim_{\substack{\varepsilon_{A},\varepsilon
_{C}\rightarrow\varepsilon\\\varepsilon_{B},\varepsilon_{D}\rightarrow
-\varepsilon}}\left.  2\frac{\partial^{2}}{\partial\varepsilon_{A}%
\partial\varepsilon_{B}}Z\right\vert _{\parallel}.\label{ZtoC}%
\end{equation}

\section{\bigskip Semiclassical generating function for orthogonal symmetry
class}

Here briefly we recapitulate the results for systems with
orthogonal symmetry. The semiclassical representation of the
generating function follows from the chain of
relations \cite{Heusl07,Muell09},%
\[
\det\left(  H-E\right)  \sim\exp\left[  -\int^{E}dE^{\prime
}\operatorname*{Tr}\left(  H-E'\right)  ^{-1}\right]
\sim\exp\left[ -\sum_{\gamma}f_{\gamma}e^{iS_{\gamma}\left(
E\right)  /\hbar}\right];
\]
 the last step is the Gutzwiller expansion of the
exponent into a sum over periodic orbits $\gamma$ with the actions $S_{\gamma
}$ and stability coefficients $f_{\gamma}$. Expanding all four exponentials we
get a sum over\ quadruplets of ``
pseudo-orbits''\ $A,B,C,D$,
\begin{align}
Z^{\left(  1\right)  }\left(  \hat{\varepsilon}\right)   & =\left\langle
\sum_{A,B,C,D}F_{A}F_{C}F_{B}^{\ast}F_{D}^{\ast}\left(  -1\right)  ^{\nu
_{C}+\nu_{D}}\right. \label{ZinS}\\
& \left.  \times e^{i\Delta S/\hbar}e^{i\left(  T_{A}\varepsilon_{A}%
+T_{C}\varepsilon_{C}-T_{B}\varepsilon_{B}-T_{D}\varepsilon_{D}\right)
/T_{H}}\right\rangle ;\nonumber
\end{align}
a pseudo-orbit, say $A,$ is a set of $\nu_{A}$  periodic orbits
whose actions and periods sum up to $S_{A},T_{A}$ and whose
product of stability coefficients is $F_{A}$. The difference of
actions $\Delta S=S_{A}+S_{C}-S_{B}-S_{D}$ must be small compared
with $\hbar$ for the quadruplets making meaningful contributions.
Note the sign factor $\left(  -1\right) ^{\nu_{C}+\nu_{D}}$
depending on the number of orbits in the pseudo-orbits associated
with the numerator of the generating function.

The leading contribution is created by the diagonal quadruplets in which the
pseudo-orbit pair $\left(  B,D\right)  $ contains the same periodic orbits as
$\left(  A,C\right)  $ and consequently $\Delta S=0\,$. \ The diagonal
contributions sum up to
\begin{equation}
Z_{\mathrm{diag}}=e^{\frac{i}{2}\left(  \varepsilon_{A}-\varepsilon
_{B}-\varepsilon_{C}+\varepsilon_{D}\right)  }\frac{\left(  \varepsilon
_{A}-\varepsilon_{D}\right)  ^{2}\left(  \varepsilon_{C}-\varepsilon
_{B}\right)  ^{2}}{\left(  \varepsilon_{A}-\varepsilon_{B}\right)  ^{2}\left(
\varepsilon_{C}-\varepsilon_{D}\right)  ^{2}}\label{Zdiag}%
\end{equation}
and can be factored out like $Z^{\left(  1\right)
}=Z_{\mathrm{diag}}\left( 1+Z_{\mathrm{off}}\right)$; the
off-diagonal part $Z_{\mathrm{off}}$ stands for a sum similar to
(\ref{ZinS}) but with all orbits of $\left( A,C\right)  $
different from those of $\left(  B,D\right)  $. Contributions with
a small action mismatch now come from the quadruplets in which the
periodic orbits of $\left(  B,D\right)  $ are `` partners''\ of
those of $\left(  A,C\right)  ,$ i.e., consist of practically the
same but differently connected pieces. Reconnections occur in ``
encounters''\ which are places of close approach of $l\geq2$
almost parallel stretches of the same or different orbits; the
possibility of such a reconnection is a fundamental property of
chaotic motion \cite{Altla09}. Using ergodicity summation over all
pseudo-orbit quadruplets could be reduced to summation over their
topological families (`` structures''). In the end
the expansion $Z_{\mathrm{off}}=\sum_{n=1}^{\infty}Z_{n},\quad$%
 was obtained \cite{Heusl07,Muell09}. Here
$Z_{n}\sim\varepsilon^{-n}$ accumulates contributions of
quadruplets with $L-V=n$; $V$ is the number of encounters
containing $L=\sum _{i=1}^{V}l_{i}$ stretches. The explicit
expression of $Z_{n}$ for the
orthogonal class is (`` O''=orthogonal),%
\begin{align}
& Z_{n,\mathrm{O}}\left(  \hat{\varepsilon}\right)  =\frac{\left(
\varepsilon_{A}-\varepsilon_{C}\right)  \left(  \varepsilon_{B}-\varepsilon
_{D}\right)  }{\left(  \varepsilon_{A}-\varepsilon_{D}\right)  \left(
\varepsilon_{B}-\varepsilon_{C}\right)  }\label{ZnGOE}\\
& \times\frac{\left(  -i\right)  ^{n}\left(  n-1\right)  !2^{n}}{\left(
\varepsilon_{A}-\varepsilon_{B}\right)  ^{n-1}}\left(  \frac{1}{\varepsilon
_{C}-\varepsilon_{D}}+\frac{n}{\varepsilon_{A}-\varepsilon_{B}}\right)
.\quad\nonumber
\end{align}

The true high-energy asymptotics of the generating function is not
exhausted by $Z^{\left(  1\right)  }$. Recovery of the missing
component was achieved on the basis of the so called ``
Riemann-Siegel look-alike'' representation
\cite{Berry90,Berry92,Keati07} of the spectral determinant, due to
the basic quantum mechanical symmetry properties. The additional
summand is obtained from $Z^{\left( 1\right) }\left(
\hat{\varepsilon}\right) $ by the ``Weyl transposition'' $w\left(
\hat{\varepsilon}\right) \equiv\left(
\varepsilon_{A},\,\varepsilon_{B,}\varepsilon_{D},\,\varepsilon
_{C}\right)  $ of its arguments, ,
\begin{align}
Z_{\mathrm{O}}\left(  \hat{\varepsilon}\right)   & \sim Z^{\left(  12\right)
}\equiv Z^{(1)}\left(  \hat{\varepsilon}\right)  +Z^{(2)}\left(
\hat{\varepsilon}\right)  ,\label{Riesig}\\
\quad Z^{(2)}\left(  \hat{\varepsilon}\right)   & =Z^{\left(  1\right)
}\left(  w\left(  \hat{\varepsilon}\right)  \right)  .
\end{align}
As shown in \cite{Muell09}, $Z^{\left(  12\right)  }$ can be
summed up and then yields the RMT generating function of the
orthogonal ensemble $Z_{\mathrm{GOE}}$.

\section{Systems with spin 1/2. Duality with spinless case}

The spin-$1/2$ evolution must be treated quantum mechanically
while the orbital motion still allows semiclassical description.
We assume that the spin is driven by the interaction with the
translational motion while neglecting the back reaction of the
spin \cite{Keppe10}. The van Vleck propagator of the two-component
wave function then falls into a product of the semiclassical
propagator of a spinless particle, and a $2\times2$ matrix of the
spin evolution. The Gutzwiller formula follows from the van Vleck
propagator after going to the energy representation and taking a
trace; consequently contribution of a periodic orbit $\gamma$ in
the Gutzwiller expansions is now to be multiplied by
$\operatorname*{Tr}U_{\gamma}$ where $U_{\gamma}$ is an $SU_{2}$
matrix describing the change of the spin state after a single
traversal of $\gamma$ \cite{Bolte99}.

The contribution of a quadruplet $(AC)(BD)$ in (\ref{ZinS}) is a
product of  Gutzwiller amplitudes of all the orbits constituting
the quadruplet. Therefore, in the presence of spin it has to be
multiplied by
the product of traces of the spin evolution matrices for all its orbits, $\Xi_{AC,BD}\equiv%
{\textstyle\prod_{\gamma\in(AC)}}
\operatorname*{Tr}U_{\gamma}%
{\textstyle\prod_{\gamma'\in(BD)}} \operatorname*{Tr}U_{\gamma'}$.
We recall that the orbits in $(AC)$ and $(BD)$ are constructed of
the same $L$ pieces. Therefore due to the group property of  the
propagator, $U_{\gamma},U_{\gamma'}$ can be replaced by products
of the matrices $\mathcal{D}_{i}$ describing the spin evolution
after traversal of the $i-$th piece. Each
$\mathcal{D}_{i},~i=1,\ldots,L$, occurs once in $(AC)$, once in
$(BD)$; if the sense of traversal of the $i-$th piece is reversed
in the partner, the second entry would be $\mathcal{D}_{i}^{-1}$.

\begin{figure}
[ptb]
\begin{center}
\includegraphics[scale=0.5,trim=0 0 0 650
]%
{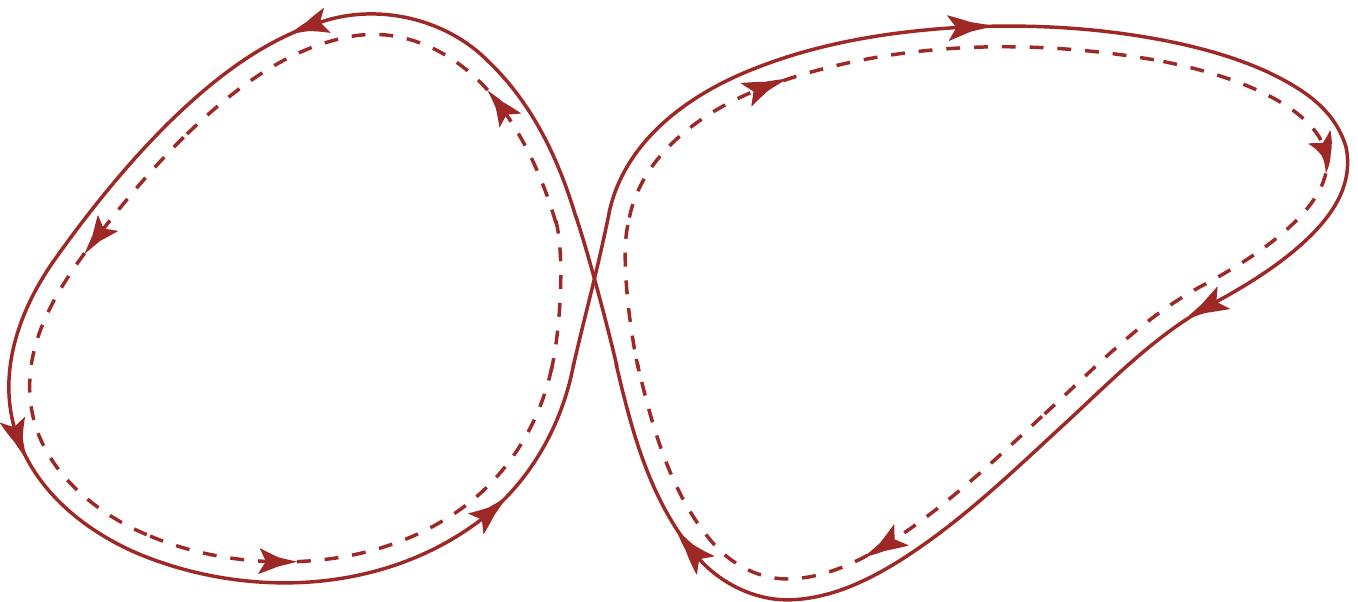}%
\caption{An orbit separates into two orbits after reconnection in
a
2-encounter}%
\label{antiSieber}%
\end{center}
\end{figure}

The factor $\Xi_{AC,BD}$ has to be averaged over an interval of
the reference energy $E$. Interaction with the chaotic
translational motion makes the spin evolution ergodic
\cite{Keppe10}. Assuming that $\mathcal{D}_{i}$ associated with
different non-overlapping pieces are independent quasi-random
$SU_{2}$ matrices, we can thus replace averaging by the
integration over all $\mathcal{D}_{i}$ over the group $SU_{2}$.
The result for the parthership of just two orbits (the only one of
interest at the sub-Heisenberg times) is well-known \cite{Heusl01,
Bolte03},
\[
\left\langle \operatorname*{Tr}U_{\gamma}\operatorname*{Tr}U_{\gamma'%
}\right\rangle =\frac{\left(  -1\right)  ^{L-V}}{2^{L-V}}%
\]
where $V,L$ are the number of encounters and encounter stretches
in the orbit pair. We generalize it to the pseudo-orbit
quadruplets containing an arbitrary number of orbits. The most
important new element is however that the orbit number can change
after reconnection in the encounters; see the elementary example
 Fig. \ref{antiSieber} where two pieces of the
figure-8 orbit become two separate orbits after reconnection.
 Calculations in
Appendix \ref{Spinfactor} show that
\begin{equation}
\left\langle \Xi_{AC,BD}\right\rangle =\frac{\left(  -1\right)  ^{L-V+\nu
_{B}+\nu_{D}-\nu_{A}-\nu_{C}}}{2^{L-V}}\label{HBHBH}%
\end{equation}
where $V,L$ are now the number of encounters and the encounter
stretches in the quadruplet $\left(  A,C\right)  ,\left(
B,D\right)  $. The additional sign factor $\left(  -1\right)
^{\nu_{B}+\nu_{D}-\nu_{A}-\nu_{C}}$ reflects the difference in the
number of orbits after the reconnection $\left(  A,C\right)
\rightarrow\left(  B,D\right)  $.

The  factor (\ref{HBHBH}) leads to important consequences. Namely,
inserting it into the semiclassical expansion (\ref{ZinS}) of the
generating function and combining with $\left( -1\right)
^{\nu_{C}+\nu_{D}} $ we obtain $\left( -1\right)
^{L-V+\nu_{B}+\nu_{D}}$. The replacement $A,C\to B,D$ in the
exponent shows that the roles of the numerator and denominator in
the generating function are reversed. We recall that a quadruplet
with $L-V=n$ contributes to $Z^{\left( 1\right)  }$ in the order
$\varepsilon ^{-n}$; our rescaling
$2\varepsilon\rightarrow\varepsilon$ in the symplectic case
absorbs $2^{L-V}$. Consequently all expansion terms of the
symplectic off-diagonal sum $Z_{\mathrm{off}}$ are obtained from
their orthogonal counterparts (\ref{ZnGOE}) by the interchange
$A\rightleftarrows C,\quad B\rightleftarrows D$ and the sign
change of all arguments,
\begin{equation}
Z_{\,n,\mathrm{S}}^{\left(  1\right)  }\left(  \hat{\varepsilon}\right)
=Z_{n,\mathrm{O}}^{\left(  1\right)  }\left(  -\varepsilon_{C},-\varepsilon
_{D},-\varepsilon_{A,}-\varepsilon_{B}\right)  .
\end{equation}
The same substitution connects the full periodic orbit expansions $Z^{\left(  1\right)  }=$ $Z_{\mathrm{diag}%
}\left(  1+\sum_{n=1}^{\infty}Z_{\,n}^{\left(  1\right)  }\right)
$ ,
\begin{equation}\label{ZSO}
Z_{\,\mathrm{S}}^{\left(  1\right)  }\left(
\hat{\varepsilon}\right) =Z_{\mathrm{O}}^{\left(  1\right)
}\left(  -\varepsilon_{C},-\varepsilon
_{D},-\varepsilon_{A,}-\varepsilon_{B}\right).
\end{equation}

 We get an important and remarkably simple relation (\ref{ZSO}) between the
semiclassical generating functions of systems with or without
half-integer spin.
 (Turning from semiclassics to RMT we note that numerous identities of that kind between the
GOE- and GSE-associated functions are well-known  under the name
of duality relations.) In Appendix \ref{Zirn} we check that a
duality relation analogous to (\ref{ZSO}) does exist between the
4-determinant generating functions of GOE and GSE. A seeming
contradiction arises: whereas the high-energy asymptotic expansion
of $Z_{\mathrm{GOE}}$ is identical with its semiclassical
counterpart $Z_{\mathrm{O}}^{\left( 12\right) }$, the analogous
expansion of $Z_{\mathrm{GSE}}$ is not: It differs from
$Z_{\,\mathrm{S}}^{\left(  12\right)  }\left(
\hat{\varepsilon}\right) =Z_{\,\mathrm{S}}^{\left(  1\right)
}\left(  \hat{\varepsilon}\right) +Z_{\,\mathrm{S}}^{\left(
1\right)  }\left(  w\left(  \hat{\varepsilon }\right)  \right)$
 by an additional elementary
summand proportional to $e^{i\left(
\varepsilon_{A}-\varepsilon_{B}\right)  /2}$, see
(\ref{SagainsGSE}). The reason is purely mathematical, and the
missing component of  the semiclassical generating function  can
be recovered by  Borel summation. We prefer to demonstrate  the
method  on the less cumbersome example of the symplectic complex
correlator, see the next Section.%

The averaged spin factor for spins different from $1/2$ is given below in (\ref{snot12}).
 For half-integer spins the result differs by the replacement of $2$
 in the denominator of (\ref{HBHBH}) by $2S+1$,
however this is compensated by the changed  mean level spacing and  Heisenberg
 time \cite{Muell05}; our equation (\ref{ZSO}) remains in
force. For integer spins the sign of the averaged spin factor is
always positive while $2S+1$ in the denominator is compensated in
the way just described, therefore the generating function is the
same as without spin.

\section{ Borel summation; missing oscillatory component as ``Stokes's satellite''}

\bigskip Applying $\partial^{2}\varepsilon_{A}\varepsilon_{B}$ to $Z^{\left(
1\right)  }\left(  \hat{\varepsilon}\right)  +Z^{\left(  2\right)
}\left( \hat{\varepsilon}\right)  $  and going to the limit in
(\ref{ZtoC}), we obtain the asymptotic expansion of the complex correlator for both symmetry classes,%
\begin{equation}
C\left(  \varepsilon\right)  \sim\sum_{n\geq2}\frac{a_{n}}{\varepsilon^{n}%
}+e^{i2\varepsilon}\sum_{n\geq4}\frac{b_{n}}{\varepsilon^{n}}%
\end{equation}
where the non-oscillatory and oscillatory parts are generated by
$Z^{(1)}\left(  \hat{\varepsilon}\right)  $ and $Z^{(2)}\left(  \hat
{\varepsilon}\right)  $ respectively. The coefficients are easily calculated
from (\ref{ZnGOE}) and the duality relation; for both symmetries $a_{2}=-1$,
while for $n>2,$
\begin{align*}
a_{n,\mathrm{O}}  & =\frac{\left(  n-3\right)  !\left(  n-1\right)  }{2\,i^{n}%
},\quad b_{n,\mathrm{O}}=\frac{\left(  n-3\right)  !\left(  n-3\right)
}{2\,i^{n}},\\
a_{n,\mathrm{S}}  & =a_{n,\mathrm{O}}\left(  -1\right)  ^{n},\quad
b_{n,\mathrm{S}}=b_{n,\mathrm{O}}.
\end{align*}
The factorial growth of the coefficients signals that the asymptotic series
diverge for all $\varepsilon$.

Suppose we want to restore the analytic functions behind these
series by means of the Borel method \cite{Balse94}. The first
stage would be the term-by-term
Fourier transform (\ref{CtoKK}) employing%
\[
\mathcal{F}\left(  \varepsilon^{-n}\right)
=\frac{\tau^{n-1}}{2i^{n}\left( n-1\right)  !}\theta\left(
\tau\right)  ,\quad\mathcal{F}\left(
e^{i2\varepsilon}\varepsilon^{-n}\right)  =\frac{\left(
\tau-2\right) ^{n-1}}{2i^{n}\left(  n-1\right)  !}\theta\left(
\tau-2\right);
\]
the resulting  series in $\tau$ and $\left(  \tau-2\right) $
converge to analytic functions. On the second stage the inverse
Fourier transform produces a closed expression for the complex
correlator. This is easily done in the orthogonal case and leads
to the form factor $K_{GOE}\left( \tau\right)  $ and then to the
exact $C_{\mathrm{GOE}}$, see Appendix \ref{RMTExplicit}, Eq.
(\ref{ExactCGOE}).

The symplectic case is more interesting. On the first stage we
obtain the form factor as
\begin{align*}
K  & =\theta\left(  \tau\right)  \left[  \tau/2-\left(  \tau/4\right)
\ln\left(  1-\tau\right)  \right] \\
& +\theta\left(  \tau-2\right)  \left[  1-\tau/2+\left(  \tau/4\right)
\ln\left(  \tau-1\right)  \right]  ;
\end{align*}
The first summand has a branch cut $(1,+\infty)$; to proceed with
Borel we need to continue $\ln\left(  1-\tau\right)  $ to all
$\tau>1$. There are three obvious choices: use the logarithm
values $\ln\left\vert 1-\tau\right\vert \pm i\pi$ at the lower or
upper lip of the cut, or their average $\ln\left\vert
1-\tau\right\vert $. Only the last option is admissible since the
form factor must be real, in view of reality of the energy
eigenvalues \cite{Haake10}; the factor at $\theta\left(
\tau\right)  $ will then be $K^{<}\equiv \tau/2-\left(
\tau/4\right)  \ln\left\vert 1-\tau\right\vert $ . The Fourier
transform of $K$ with this choice produces the \emph{exact} GSE
correlator (\ref{ExactCGSE}) whose asymptotic expansion contains
an additional oscillatory summand,
\[
C_{GSE}\sim\sum_{n=2}^{\infty}\frac{a_{n,S}}{\varepsilon^{n}}+e^{2i\varepsilon
}\sum_{n=2}^{\infty}\frac{b_{n,S}}{\varepsilon^{n}}+\frac{\pi}{2\varepsilon
^{2}}\left(  \varepsilon+i\right)  e^{i\varepsilon};
\]
incidently, the latter would be generated by the term $\propto e^{i\left(
\varepsilon_{A}-\varepsilon_{B}\right)  /2}$ of $Z_{\mathrm{GSE}}$
(\ref{SagainsGSE}) after application of (\ref{ZtoC}).

It may seem strange that the back-and-forth Fourier transform
recovered, free of charge, the missing oscillatory contribution to
the correlator. In fact, restoration of an oscillatory term, given
an asymptotic power series, is a legitimate mathematical tool
described in detail in the book \cite{Dingl73}; see Berry and
coauthors \cite{Berry89,Berry94} for further important
developments. The key idea is that the manner in which
coefficients of an asymptotic series tend to infinity contains
information about the exponentially small terms disregarded in the
classical Poincar\'{e} approach. Such terms become oscillatory and
of crucial importance when the asymptotics is continued to the
anti-Stokes lines.

\begin{figure}
[ptb]
\begin{center}
\includegraphics[scale=0.3,trim=0 150 0 150]
{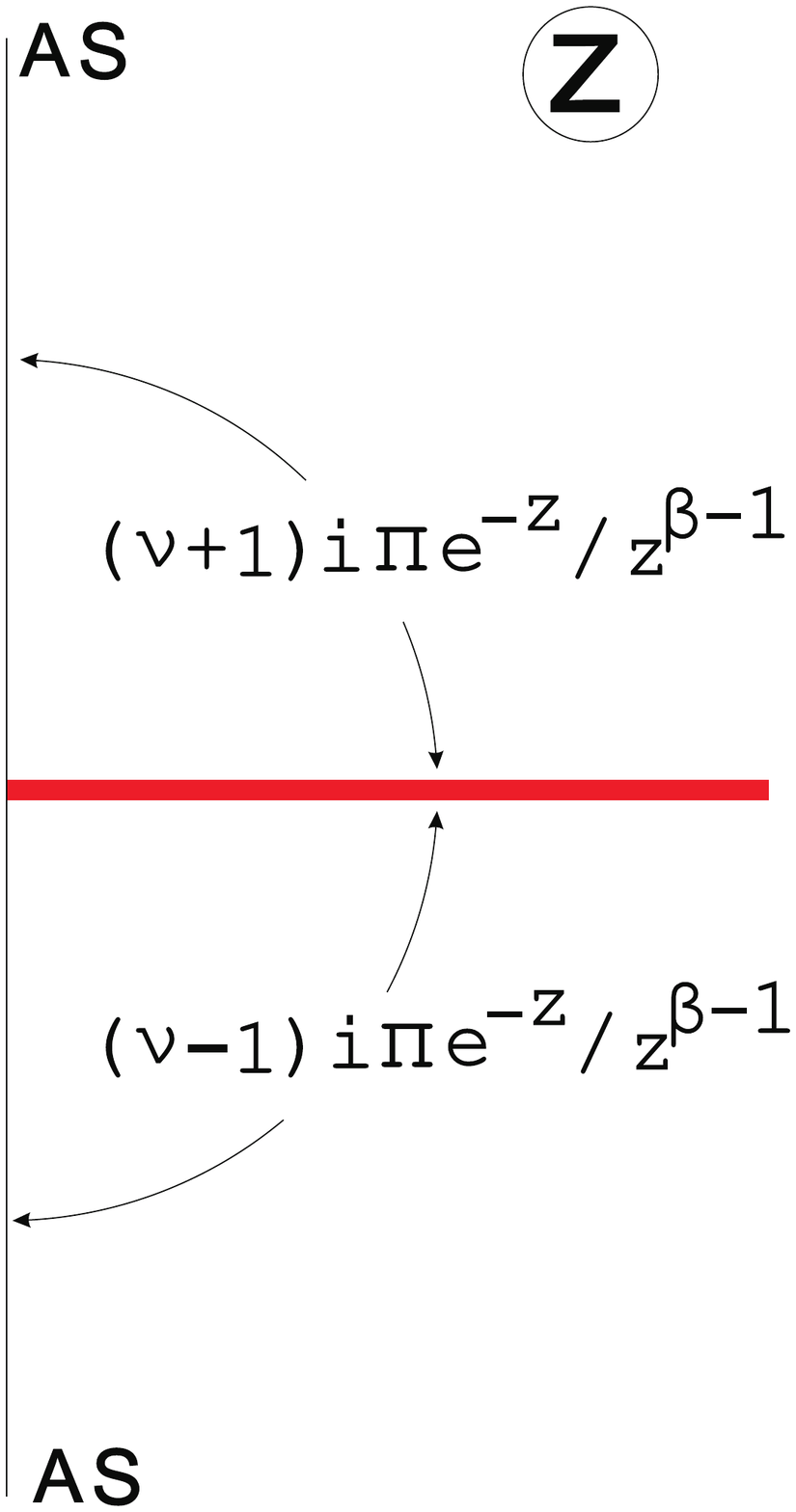}%
\caption{The exponential ``satellite'' changes by a jump at the
Stokes
 line (red) and becomes oscillatory at the anti-Stokes (AS) lines}%
\label{figDingle}%
\end{center}
\end{figure}

Here is the barest minimum of detail on the method. Consider a diverging
asymptotic expansion of an analytic function $g\left(  z\right)  ,\quad
z=x+iy, $%
\begin{equation}
g\left(  z\right)  \sim\sum_{n=1}^{\infty}\frac{c_{n}}{z^{n}},\quad\left\vert
z\right\vert \rightarrow\infty,\label{series}%
\end{equation}
and assume that in the limit of large $n$\ \ its coefficients tend to%
\begin{equation}
c_{n}\rightarrow\left(  n-\beta\right)  !
\end{equation}
Then

a) The real positive semi-axis is the Stokes line at which the
power expansion (\ref{series}) has all its terms positive and
therefore maximally dominant with respect to an exponentially
small additional  component $g_{SD}\left( z\right)  $; the value
of $g_{SD}\left(  z\right)$ changes almost by a jump when the
positive semi-axis is crossed;

b) The subdominant component behaves like%
\[
g_{SD}\left(  z\right)  \propto\frac{e^{-z}}{z^{\beta-1}};
\]

c) The imaginary semiaxes are the anti-Stokes lines where
$g_{SD}\left( z\right)  $ becomes oscillatory and comparable to
the power expansion, see Fig.~\ref{figDingle};

d) Under certain assumptions about the properties of $g(z)$ we have,
\begin{align}
g\left(  z\right)   & =\sum_{n=1}^{n^{\ast}\left(  z,q\right)  }\frac{c_{n}%
}{z^{n}}+R\left(  z\right)  ,\label{Dingle}\\
R\left(  z\right)   & \approx e^{-z}\frac{\pi}{z^{\beta-1}}\left[
i\operatorname{erf}\left(  \frac{y}{\sqrt{2x}}\right)  +i\,\nu+\eta\left(
z,q\right)  \right]  .
\end{align}
The upper limit of the sum is $n^{\ast}\left(  z,q\right)  =\operatorname{Int}%
\left\{  \left\vert z\right\vert +q\right\}  $ where $q$ is of the
order unity and otherwise arbitrary. The error function
$\operatorname{erf}\left( \sigma\right)
=\frac{2}{\sqrt{\pi}}\int_{0}^{\sigma}e^{-t^{2}}dt$ in
(\ref{Dingle}) is close to $1$  for $\left\vert z\right\vert $
large and $\arg z>\delta>0,$ and to $-1$ for $\arg z<-\delta<0$;
on the real axis it is zero. The almost jump-like change of the
subdominant component when the real axis is crossed, is the Stokes
phenomenon. The small real correction
\[
\eta\left(  z,q\right)  =\frac{2}{\sqrt{2\pi x}}\left(  \operatorname{Fract}%
\left\{  \left\vert z\right\vert +q\right\}  +\beta-q-\frac{4}{3}-\frac{y^{2}%
}{6x}\right)  e^{-y^{2}/2x}.
\]
is significantly non-zero only close to the $x-$axis; its dependence on $q$
compensates that of the sum in (\ref{Dingle}) . The constant $\nu$ must be
deduced from additional information on the function $g\left(  z\right)  $. In
particular, if $g\left(  z\right)  $ is real on the real axis we must choose
$\nu=0$; the asymptotics of $g\left(  z\right)  $ contains then oscillatory
components at both anti-Stokes lines.

Let us apply the method to the non-oscillatory part of the symplectic
correlator. The terms of its expansion $a_{n,\mathrm{S}}/\varepsilon^{n}$ are
all positive at the positive imaginary axis of $\varepsilon$ which is
 the Stokes line where the power series is maximally dominant; the
anti-Stokes lines are the real semiaxes of $\varepsilon$. The
results above are applicable with $\varepsilon=iz,\ \,\beta=2$ .
According to its definition (\ref{defcorr}), the complex
correlator must be real for positive imaginary $\varepsilon$.
Therefore we must choose in (\ref{Dingle}) $\nu=0$ such that the
oscillatory `` Stokes satellite''\
of the power series must be present on both real semiaxes,
\begin{align}
R\left(  \varepsilon\right)   & \approx e^{i\varepsilon}\frac{\pi
}{2\varepsilon},\quad\varepsilon\rightarrow+\infty,\,\,\,\label{leadingorder}%
\\
R\left(  \varepsilon\right)   & \approx-e^{i\varepsilon}\frac{\pi
}{2\varepsilon}=e^{i\varepsilon}\frac{\pi}{2\left\vert \varepsilon\right\vert
},\quad\varepsilon\rightarrow-\infty.
\end{align}
This is indeed the leading term in the oscillatory part of the symplectic
correlator recovered by the Borel method. It is subdominant in the upper
half-plane away from the real axis and experiences the $\operatorname{erf}%
-$like approximate discontinuity at the positive imaginary axis.

It is instructive to investigate what  happens if we choose to
continue $\log\left(  1-\tau\right)  \ $\ as $\log\left(
\tau-1\right)  \pm i\pi$ for $\tau>1$ in the form factor at the
first stage of the Borel summation. The functions obtained by the
inverse Fourier transform would then differ from the correct
complex correlator by the additional terms
\[
\mp\frac{i\pi}{2}\int_{1}^{\infty}d\tau e^{i\varepsilon\tau}\tau=\pm\frac{\pi
}{2\varepsilon^{2}}\left(  i+\varepsilon\right)  e^{i\varepsilon}\sim\pm
e^{i\varepsilon}\frac{\pi}{2\varepsilon}%
\]
They would cancel (\ref{leadingorder}) at one of the real semiaxes and double
its amplitude at the other one, i.e., exactly what we would get if we chose
$\nu=\pm1$ in (\ref{Dingle}). Therefore the alternative choices of $\nu$ are
equivalent to different continuation of the result of the first Borel stage
beyond the branch point in the $\tau$ domain.

Finally let us convince ourselves that additional oscillatory
components do not arise in the orthogonal case. The
$\varepsilon^{-1}-$expansion with the coefficients
$a_{n,\mathrm{O}}$ has its terms all positive on the negative
imaginary semi-axis, i.e. in the non-physical half-plane of
$\varepsilon$ where (\ref{defcorr}) is inapplicable and the
correlator need not be real. The parameter $\beta=2$ is the same
as in the symplectic case but we must now set $\varepsilon=-iz$
and the Stokes satellite now behaves like $\sim
e^{-i\varepsilon}/\varepsilon$. Let us change the phase of
$\varepsilon$ from $-\pi/2$ via $0$ to positive values; if the
satellite were present with a non-zero amplitude it would become
exponentially large in the physical region
$\operatorname{Im}\varepsilon>0$. This is forbidden, and we must
choose thus in (\ref{Dingle}) $\nu=-1$ which corresponds to
absence of the exponential term in the sector $-\pi
/2+\delta<\arg\varepsilon<\pi$, in particular at the real positive
semi-axis.

\section{\bigskip Conclusion}

We studied the generating function, complex correlator and  form
factor of systems with  spin $1/2$. Expanding the generating
function into a sum over periodic orbit quadruplets we showed that
in the presence of spin, the terms of the expansion acquire an
additional sign factor whose effect is to interchange the role of
the numerator and the denominator of the generating function. As a
result, the generating functions of the orthogonal and symplectic
class turn out to be connected by a simple substitution of their
arguments.

The periodic orbit expansion of the generating function
supplemented by the  Riemann-Siegel look-alike formula for the
spectral determinants yields the complex correlator  as
combination of two asymptotic
series in $\varepsilon^{-1}$, the second one multiplied by $e^{i2\varepsilon}%
$; they are responsible for the form factor at small times and at
times larger than the Heisenberg time. We demonstrate how the
Borel summation reveals in the symplectic case one more
oscillatory term $\propto e^{i\varepsilon}$ associated with the
logarithmic singularity of the form factor; the origin of that
term is clarified by the Dingle-Berry method of smart summation of
the asymptotic series as a display of the Stokes phenomenon. With
the missing oscillatory term restored, complete equivalence of the
correlation functions of  chaotic systems with  spin $1/2$ and the
Gaussian symplectic ensemble of RMT is reached.

The inherent ambiguity in restoration of a function from its asymptotic series
is solved on the ground of reality of the energy eigenvalues. The same reason
is at the heart of the Riemann-Siegel look-alike, such that existence of both
oscillatory components of the complex correlator in the symplectic case can be
traced to unitarity of the quantum mechanical evolution.

There are several possible further developments of the theory. An
obvious generalization would be to consider parametric correlation
in systems with spin $1/2$ at times comparable with the Heisenberg
time \cite{Nagao07}. Away from the deep semiclassical limit,
system-specific deviations from the universal behavior in systems
with half-integer spin can be of physical interest.

\section{\bigskip Acknowledgement}

I want to thank  Fritz Haake for continuous help and support in
the years of my work in Essen and Duisburg; he read the paper
before publication and made many important suggestions. The author
is grateful to  Martin Zirnbauer for providing his unpublished
results on the generating functions of RMT, and Stefan Heusler,
Sebastian Mueller,  Thomas Guhr and Yan Fedorov for useful
discussions. Financing by Sonderforschungsbereich TR12 is
acknowledged.

\section{Appendix}

\subsection{RMT complex correlator of the orthogonal and symplectic case
\label{RMTExplicit}}

\bigskip The  complex correlators  are  conveniently
expressed in terms of the  functions,%
\[
f_{\pm}(z)=\int_{z}^{\infty}\frac{e^{\pm i\left(  t-z\right)
}}{t}dt;
\]
The integral representations are applicable whenever the integral
converges; for all $z$ we have
\begin{equation}\label{deffplusminus}
f_{\pm}\left(  z\right)  \equiv e^{\mp iz}\left[  \pm i\frac{\pi}%
{2}-\operatorname{Ci}\left(  z\right)  \mp i\operatorname{Si}\left(  z\right)
\right]  .
\end{equation}
where $\operatorname{Ci}$, $\operatorname{Si}$ are the integral
sine and cosine. These functions are analytic in the plane of $z$
with the cut along the negative real axis.

The complex correlator of the orthogonal case can be represented in terms of
$f_{\pm}$ as,%
\begin{align}
C_{\mathrm{GOE}}\left(  \varepsilon\right)   & =2\int_{0}^{\infty
}e^{i2\varepsilon\tau}K_{\mathrm{GOE}}\left(  \tau\right)  d\tau
=C_{\mathrm{GOE}}^{\left(  1\right)  }+C_{\mathrm{GOE}}^{\left(  2\right)
},\nonumber\\
C_{\mathrm{GOE}}^{\left(  1\right)  }  & =-\frac{1}{2\varepsilon^{2}}+\frac
{1}{2\varepsilon^{2}}\left(  i\varepsilon+1\right)  f_{+}(\varepsilon
),\nonumber\\
C_{\mathrm{GOE}}^{\left(  2\right)  }  & =-\frac{e^{i2\varepsilon}%
}{2\varepsilon^{2}}\left[  \left(  -i\varepsilon+1\right)  f_{+}%
(\varepsilon)-1\right]  ,\label{ExactCGOE}%
\end{align}
while in the symplectic case
\begin{align}
C_{\mathrm{GSE}}\left(  \varepsilon\right)   & =2\int_{0}^{\infty
}e^{i\varepsilon\tau}K_{\mathrm{GSE}}\left(  \tau\right)  d\tau
=C_{\mathrm{GSE}}^{\left(  1\right)  }+C_{\mathrm{GSE}}^{\left(  2\right)
}+C_{\mathrm{GSE}}^{\left(  3\right)  },\nonumber\\
C_{\mathrm{GSE}}^{\left(  1\right)  }  & =-\frac{1}{2\varepsilon^{2}}+\frac
{1}{2\varepsilon^{2}}\left(  -i\varepsilon+1\right)  f_{-}(\varepsilon
),\nonumber\\
C_{\mathrm{GSE}}^{\left(  2\right)  }  & =-\frac{e^{2i\varepsilon}%
}{2\varepsilon^{2}}\left[  \left(  -i\varepsilon+1\right)  f_{+}%
(\varepsilon)-1\right]  .\nonumber\\
C_{\mathrm{GSE}}^{\left(  3\right)  }  & =\frac{\pi}{2\varepsilon^{2}}\left(
\varepsilon+i\right)  e^{i\varepsilon}.\label{ExactCGSE}%
\end{align}

\begin{figure}
[ptb]
\begin{center}
\includegraphics[scale=0.25]
{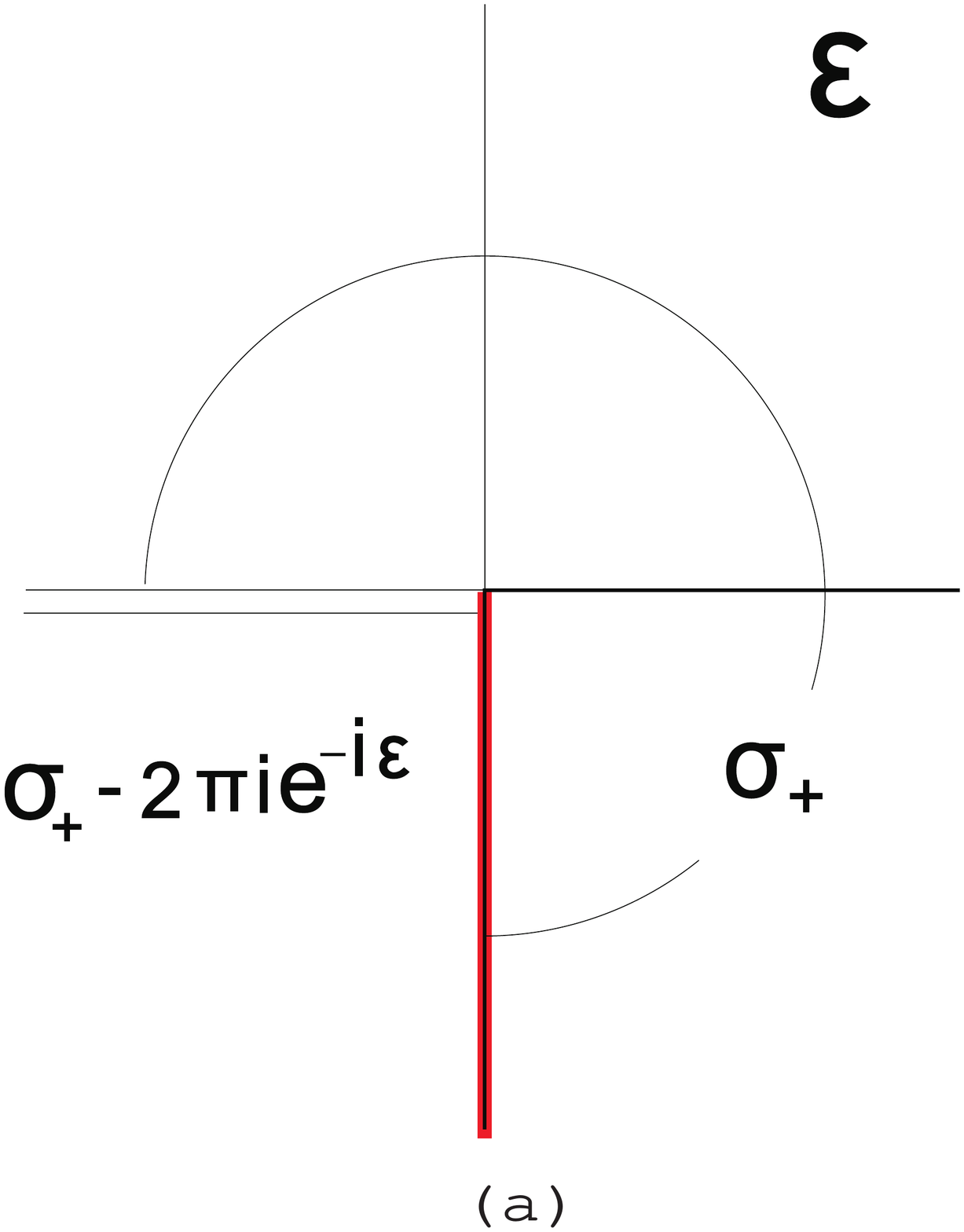}%
\hspace{0.5cm}
\includegraphics[scale=0.25]
{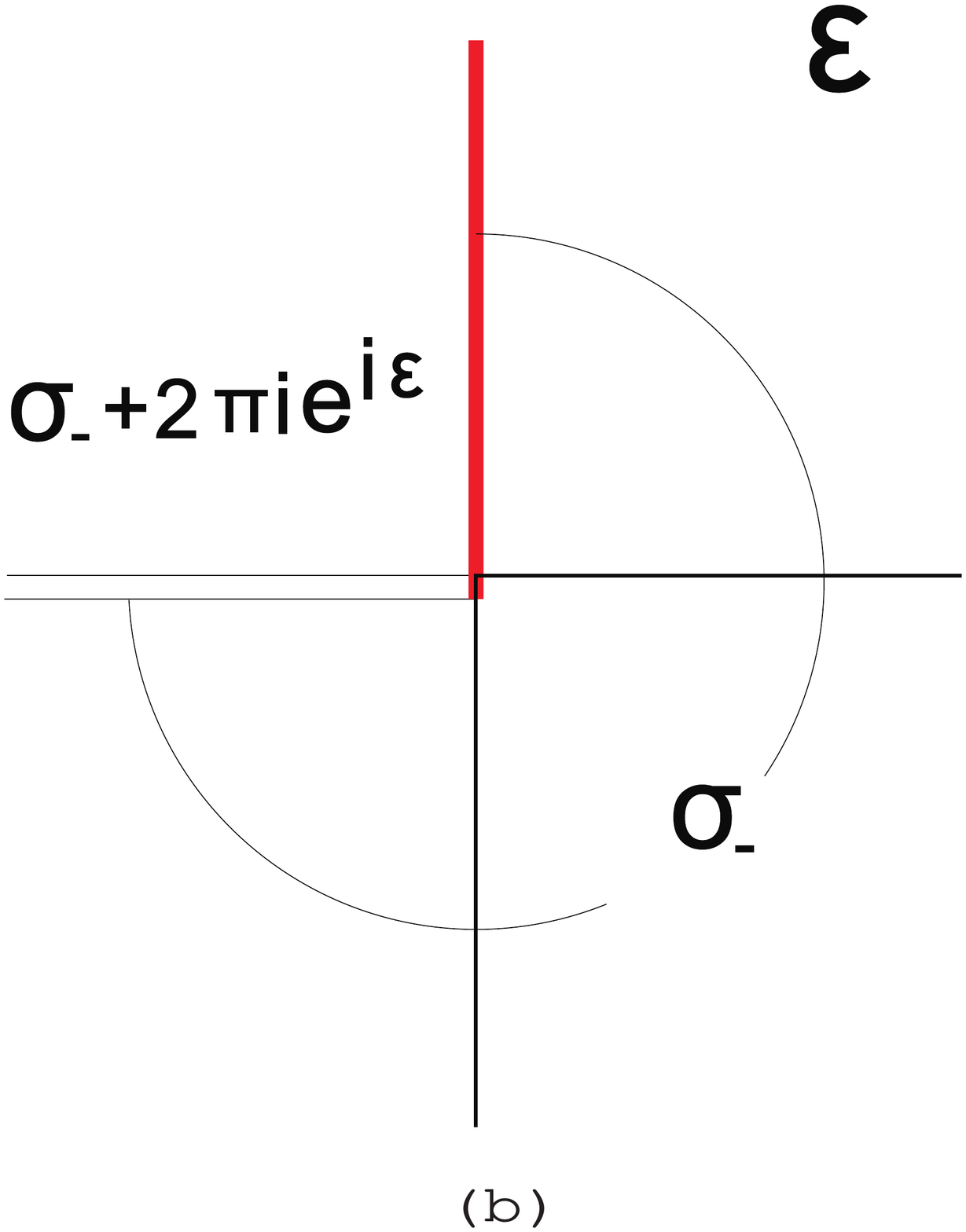}%
\caption{Asymptotics of the functions $f_{+}\left(
\varepsilon\right)  $ (a) and $f_{-}\left( \varepsilon\right)$
(b): exponential term is present left of the Stokes line  (red)}%
\label{fplusminus}%
\end{center}
\end{figure}

The components $C^{\left(  1\right)  }$ in both cases have non-oscillatory
asymptotic expansion in powers of $\varepsilon^{-1}$ at the real positive
axis~while $C^{\left(  2\right)  }$ oscillates like $e^{i2\varepsilon}$. That
follows from the asymptotic representations of $f_{\pm}\left(  z\right)  $,
\begin{align}
f_{\pm}\left(  z\right)    & \sim-\sum_{k=0}^{\infty}\frac{k!}{\left(  \pm
iz\right)  ^{k+1}}\equiv\sigma_{\pm}\left(  z\right)  ,\quad|z|\rightarrow
\infty,\label{sigmadef}\\
-\pi+\delta & <\arg z\leq\pi-0,\quad\quad(f_{+});\nonumber\\
-\pi+0  & \leq\arg z<\pi-\delta.\quad\quad(f_{-}).\nonumber
\end{align}
The Stokes line where the power expansion is dominant is
$z=-it,\,\,\,t>0,$ for $f_{+}\left( z\right) $ and
$z=it,\,\,\,t>0,$ for $f_{-}\left( z\right) $; the subdominant
satellite, $-2\pi ie^{-iz}$ for $\sigma_{+}\left( z\right)  ~$and
$2\pi ie^{iz}$ for $\sigma_{-}\left(  z\right)  $, exists in the
quadrant to the left
 of the respective Stokes line, see Fig.~\ref{fplusminus}.

$\quad$The form factors obtained from the complex correlators by the
transformation (\ref{CtoKK}) are

\begin{itemize}
\item GOE:%
\begin{align*}
K\left(  \tau\right)   & =2\tau-\tau\ln\left(  1+2\tau\right)  ,\quad\tau<1;\\
K\left(  \tau\right)   & =2-\tau\ln\frac{2\tau+1}{2\tau-1},\quad\tau>1.
\end{align*}

\item GSE:%
\begin{align*}
K\left(  \tau\right)   & =\frac{\tau}{2}-\frac{\tau}{4}\ln\left\vert
1-\tau\right\vert ,\quad\tau<2;\\
K\left(  \tau\right)   & =1,\quad\tau>2.
\end{align*}

\end{itemize}

\subsection{Spin factor for quadruplets\label{Spinfactor}}

Contribution of the quadruplet $(AC)(BD)$ in the periodic orbit expansion of
the generating function (\ref{defZ}) contains in the symplectic case the spin
factor $\Xi_{AC,BD}\equiv$ $%
{\textstyle\prod_{\gamma\in(AC)}}
\operatorname*{Tr}U_{\gamma}%
{\textstyle\prod_{\gamma'\in(BD)}} \operatorname*{Tr}U_{\gamma'}$
which is product of traces of the periodic orbits composing the
quadruplet. Non-vanishing contributions to the sum are created by
the quadruplets such that the orbits of $(BD)$ are constructed
from pieces of the orbits in $(AC)$ connected in different order
and possibly traversed with a different sense. We assume
ergodicity of the spin motion and independence of the spin
evolution along different orbit pieces. Averaging is then done
step by step by integration over the evolution matrices
$\mathcal{D}_{l}$ associated with the orbit pieces; the
integration domain is the group $SU_{2}$. Possible outcomes of a
single integration are
summed up by the relations \cite{Bolte03},%

\begin{align}
\int d\mathcal{D}\operatorname*{Tr}\left(  \mathcal{ADBD}\right)   &
=-\frac{1}{2}\operatorname*{Tr}\left(  \mathcal{AB}^{-1}\right)
,\label{recur1}\\
\int d\mathcal{D}\operatorname*{Tr}\left(  \mathcal{ADBD}^{-1}\right)   &
=\frac{1}{2}\operatorname*{Tr}\mathcal{A}\operatorname*{Tr}\mathcal{B}%
,\label{recur2}\\
\int d\mathcal{D}\operatorname*{Tr}\left(  \mathcal{AD}\right)
\operatorname*{Tr}\left(  \mathcal{BD}\right)   & =\frac{1}{2}%
\operatorname*{Tr}\left(  \mathcal{AB}^{-1}\right)  .\label{recur3}%
\end{align}
Here $\mathcal{A},\mathcal{B}$ are any fixed $SU_{2}$ matrices.
Applying these rules, e. g., to the structure with $V=1,L=2$ shown
in Fig. \ref{antiSieber}, we obtain the result,
\begin{eqnarray*}
\langle\Xi_{AC,BD}\rangle=\int d\mathcal{D}_1d\mathcal{D}_2
\operatorname*{Tr}\left(
\mathcal{D}_1\mathcal{D}_2\right)\operatorname*{Tr}\left(
\mathcal{D}_1\right) \operatorname*{Tr}\left(
\mathcal{D}_2\right)=\frac 1 2,
\end{eqnarray*}
which is opposite in sign compared with the Sieber-Richter pair
 \cite{Heusl01}.

We shall find the average of $\Xi_{AC,BD}$ for all structures
extending the inductive method of Bolte and Harrison from pairs of
orbits to the pseudo-orbits quadruplets. We remind the essence of
the method. Consider an orbit pair $\gamma ,\gamma'$ differing in
$V$ encounters with $L$ encounter stretches and assume that
averaging of the spin factor produces the factor $C_{\gamma
\gamma'}=\left(  -1\right) ^{L-V}/2^{L-V}$. Introduce one more
orbit $\gamma''$ differing from $\gamma'$ by an additional
$2-$encounter such that the pair $\gamma,\gamma''$ contains
$V''=V+1$ active encounters with $L''=L+2$ stretches. It is then
shown using the recurrence relations that
$C_{\gamma\gamma''}=-C_{\gamma\gamma'}/2=\left( -1\right)
^{L''-V''}/2^{L''-V''}$. Starting from a pair without active
encounters $L=V=0$ when the formula is true, and adding
$2-$encounters one by one we obtain that the result is true for an
arbitrary number of $2-$encounters. Finally, reconnection in any
$l-$encounter with $l>2$ can be reduced to $l-1$ successive
reconnections in $2-$encounters in $l-1~$steps, with the factor
$\left(  -1\right) ^{L-V}/2^{L-V}$  correct on all steps. Indeed,
reconnection in an $l-$encounter can be described by a permutation
of $l$ elements, however, any permutation can be represented as a
chain of transpositions of just two elements.

Let us apply this method to the pseudo-orbit quadruplets. Let
$\Gamma=(A,C)$ be the initial pseudo-orbit pair, and
$\Gamma'=\left(  B,D\right)  $ its partner differing from $\Gamma$
in $V$ active encounters with $L$ stretches. Let $\Gamma''$ be a
pseudo-orbit pair differing from $\Gamma'$ by reconnection in a
single additional $2-$encounter whose
stretches belong to some orbit $\gamma$ in $\Gamma$ and to $\gamma'%
\ $\ in $\Gamma'$ where $\gamma,\gamma'$ may differ in an
arbitrary number of other encounters. The only new case to be
considered is that of a $2-$encounter with almost parallel
stretches such that
$\gamma'$ breaks up after reconnection into two orbits $\gamma_{p}''%
$ and $\gamma_{q}''$ belonging to $\Gamma'' $, see the example in
Fig.~\ref{antiSieber}.

The averaged spin factors for the quadruplets $\Gamma\Gamma'$ and
$\Gamma\Gamma''$can be written as multiple integrals over
$SU_{2}$ with the Haar measure,%
\begin{align*}
C_{\Gamma\Gamma'}  & =\int d\left(  \ldots\right)  \int d\left(
\gamma\right)  d\left(  \gamma'\right)  ,\\
C_{\Gamma\Gamma''}  & =\int d\left(  \ldots\right)  \int d\left(
\gamma\right)  d\left(  \gamma_{p}''\right)  \left(
\gamma_{q}''\right)
\end{align*}
Here $d\left(  \gamma\right)  ,d\left(  \gamma'\right)  $ denote
$SU_{2}$ integration over matrices associated with pieces of
$\gamma$ and
$\gamma'$ while $ d\left(  \gamma_{p}''%
\right)  d\left(  \gamma_{q}''\right)  $ indicate integration over
matrices associated with pieces of the disconnected orbit pair$\
$\ in $\Gamma''$. Integration $d\left( \ldots\right)  $ is over
the remaining variables, same for $\Gamma\Gamma'$ and
$\Gamma\Gamma''$. In the way of induction, let us assume that
\begin{equation}
C_{\Gamma\Gamma'}=\frac{\left(  -1\right)  ^{L-V+\nu_{\Gamma}%
-\nu_{\Gamma'}}}{2^{L-V}}\label{assumption}
\end{equation}
where $\nu_{\Gamma}=\nu_{A}+\nu_{C}$ and $\nu_{\Gamma'}=\nu_{B}%
+\nu_{D}$ is the number of periodic orbits in\ $\Gamma$ and
$\Gamma^{\prime }$, respectively, and prove that the analogous
formula will be true for $C_{\Gamma\Gamma'' }$.

\begin{figure}[ptb]
\begin{center}
\includegraphics[scale=0.6,angle=-90]{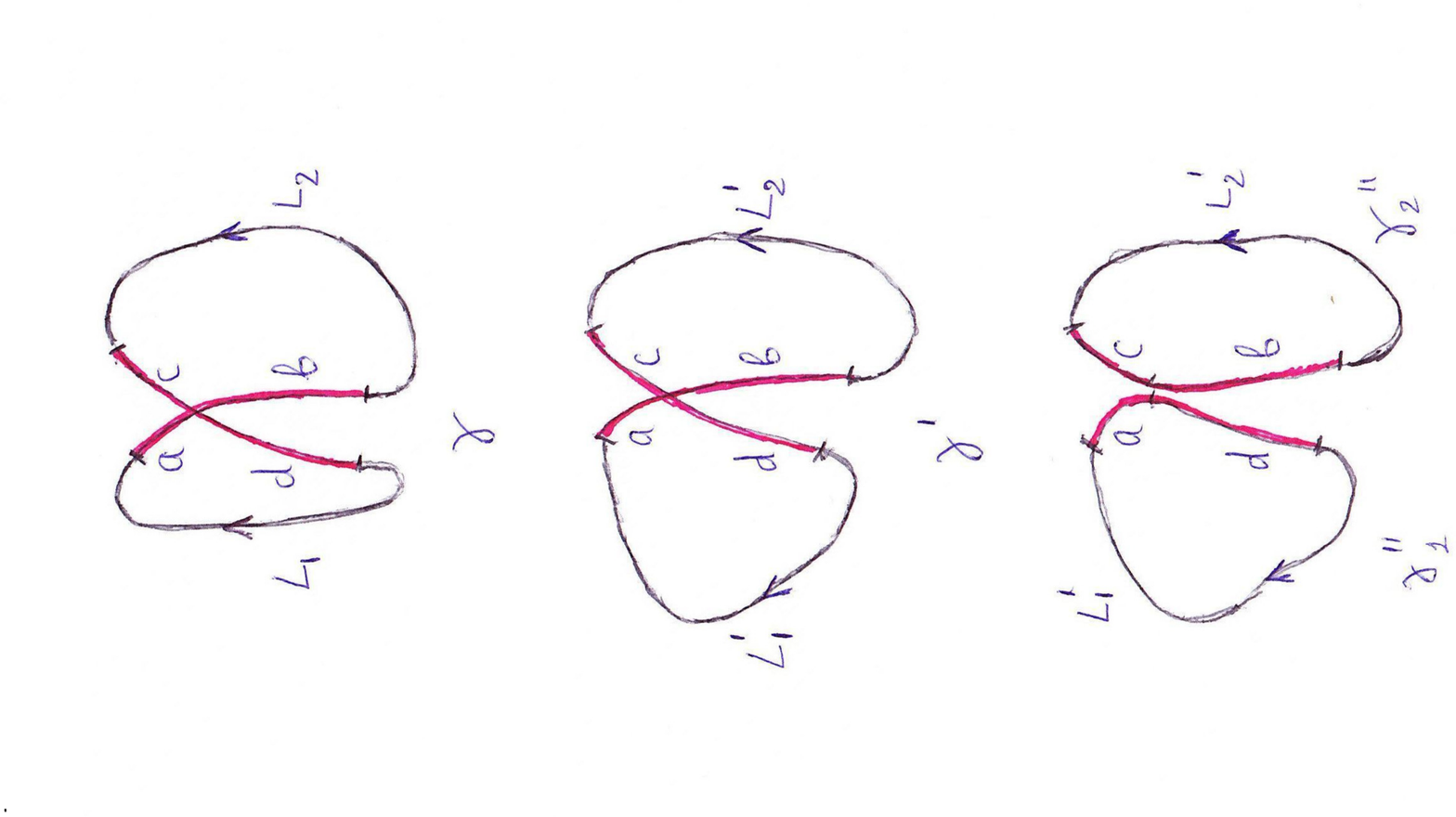}%
 \caption{A 2-encounter belongs to the orbits $\gamma$ in the pseudo-orbit $\Gamma$ and
 $\gamma' $ in the partner pseudo-orbit
$\Gamma'$. Reconnection in that 2-encounter divides $\gamma'$ into
$\gamma_{1}'',\gamma _{2}''$  creating
a new pseudo-orbit $\Gamma''$}%
\label{BolteHar}
\end{center}
\end{figure}
Consider Fig. \ref{BolteHar} where the orbits $\gamma,\gamma'$ and
the orbit pair $\gamma''$ are depicted. In $\gamma,\gamma'$ we see
a parallel crossing; however it is not switched between $\gamma
,\gamma'$, i.e., it is inactive and not counted in
$V=V_{\Gamma\Gamma'}$. On the other hand, it is activated in the
pair $\gamma\gamma''$ such that
\[
V_{\Gamma\Gamma''}=V+1,\quad L_{\Gamma\Gamma''}=L+2,\quad\nu_{\Gamma''}%
=\nu_{\Gamma'}+1.
\]
In Fig. \ref{BolteHar} the orbit pieces adjacent to the encounter
and incorporating parts of its stretches are denoted $a,b,c,d$;
they are assumed to coincide in $\gamma,\gamma'$, i.e., do not
contain any additional active encounters. The associated matrices
$\mathcal{D}_{a}$\ etc. involved in the $SU_{2}$ integration will
be denoted by the same letters for compactness,
$a\equiv\mathcal{D}_{a}$ etc. (we hope that $d$ as the integration
variable will not be mixed with $d$ as the differential!) The two
links attached to the selected
$2-$encounter are denoted $L_{1},L_{2}$ in $\gamma$ and $L_{1}^{'%
},L_{2}^{'}$ in $\gamma'$. Unlike $a,b,c,d,$ the links $L_{i\text{
}}$and $L_{i}'$ need not coincide; indeed, $L_{i}$ can contain any
amount of encounters active in $\Gamma\Gamma'$ such that $L_{i}'$
can contain pieces of all orbits of the original quasi-orbit
pair $\Gamma$ other than $\gamma$.%

The $SU_{2}$ integrals can be written, with $dL_{1}$ being a
shorthand for
$d\mathcal{D}_{L_{1}},$ etc,%
\begin{align*}
C_{\Gamma\Gamma'}  & =\int d\left(  \ldots\right)  dL_{1}dL_{2}%
dL_{1}'dL_{2}'\int da\,db\,dc\,dd\,\operatorname*{Tr}\left(
L_{1}abL_{2}cd\right)  \operatorname*{Tr}\left(  L_{1}'abL_{2}%
'cd\right)  ;\\
C_{\Gamma\Gamma''}  & =\int d\left(  \ldots\right)
dL_{1}dL_{2}dL_{1}^{\prime
}dL_{2}'\int da\,db\,dc\,dd\,\operatorname*{Tr}\left(  L_{1}%
abL_{2}cd\right)  \operatorname*{Tr}\left(  L_{1}'ad\right)
\operatorname*{Tr}\left(  L_{2}'cb\right)  .
\end{align*}
Let us transform the integrals over $a,b,c,d$. In
$C_{\Gamma\Gamma'}$ we can take the matrices $x=ab$ and $y=cd$ as
the new integration variables writing
\begin{align}
& \int da\,db\,dc\,dd\,\operatorname*{Tr}\left(
L_{1}abL_{2}cd\right) \operatorname*{Tr}\left(
L_{1}'abL_{2}'cd\right)
\label{GammaGammaprim}\\
& =\int dx\,dy\operatorname*{Tr}\left(  L_{1}xL_{2}y\right)
\operatorname*{Tr}\left(  L_{1}'xL_{2}'y\right) \nonumber\\
& =\frac{1}{2}\int dy\operatorname*{Tr}\left[  L_{2}yL_{1}\left(
L_{1}^{'}\right)  ^{-1}y^{-1}\left(  L_{2}^{'}\right)
^{-1}\right] \nonumber\\
& =\frac{1}{4}\operatorname*{Tr}\left[  L_{1}\left(
L_{1}^{'}\right) ^{-1}\right]  \operatorname*{Tr}\left[
L_{2}\left(  L_{2}^{'}\right) ^{-1}\right]  ;\nonumber
\end{align}
The cyclic invariance of the trace and the relations (\ref{recur3}) and
(\ref{recur2}) were used.

Now let us carry out similar transformations of the integral in
$C_{\Gamma \Gamma''}$ introducing consecutively new integration
variables
$x=ad$,$\,\,\ y=cb,z=cd$,%
\begin{align*}
& \int da\,db\,dc\,dd\,\operatorname*{Tr}\left(
L_{1}abL_{2}cd\right) \operatorname*{Tr}\left(  L_{1}'ad\right)
\operatorname*{Tr}\left(
L_{2}'cb\right) \\
& =\int dx\,db\,dc\,dd\,\operatorname*{Tr}\left(
L_{1}xd^{-1}bL_{2}cd\right) \operatorname*{Tr}\left(
L_{1}'x\right)  \operatorname*{Tr}\left(
L_{2}'cb\right) \\
& =\frac{1}{2}\int\,db\,dc\,dd\,\operatorname*{Tr}\left[  d^{-1}bL_{2}%
cdL_{1}\left(  L_{1}'\right)  ^{-1}\right]
\operatorname*{Tr}\left(
L_{2}'cb\right) \\
& =\frac{1}{2}\int dy\,dc\,dd\,\operatorname*{Tr}\left[  d^{-1}c^{-1}%
yL_{2}cdL_{1}\left(  L_{1}'\right)  ^{-1}\right]  \operatorname*{Tr}%
\left(  L_{2}'y\right) \\
& =\frac{1}{4}\int\,dc\,dd\,\operatorname*{Tr}\left[
L_{2}c\,d\,L_{1}\left( L_{1}'\right)  ^{-1}d^{-1}c^{-1}\left(
L_{2}'\right)
^{-1}\right] \\
& =\frac{1}{4}\int dz\operatorname*{Tr}\left[  \,L_{1}\left(  L_{1}^{\prime
}\right)  ^{-1}z\left(  L_{2}'\right)  ^{-1}L_{2}z^{-1}\right] \\
& =\frac{1}{8}\operatorname*{Tr}\left[  L_{1}\left(
L_{1}^{'}\right) ^{-1}\right]  \operatorname*{Tr}\left[
L_{2}\left(  L_{2}^{'}\right) ^{-1}\right]  .
\end{align*}
Comparing with (\ref{GammaGammaprim}) we see that,%
\[
C_{\Gamma\Gamma''}=\frac{1}{2}C_{\Gamma\Gamma'}%
\]
which agrees with (\ref{assumption}) with $L\rightarrow
L^{\prime\prime }=L+2,V\rightarrow
V''=V+1,\nu_{\Gamma'}\rightarrow \nu_{\Gamma''}=\nu_{\Gamma'}+1$ ;
therefore if (\ref{assumption}) is true for the quadruplet
$\Gamma\Gamma'$ it will also hold for $\Gamma\Gamma''$.

Evidently, the reversed process when two orbits of $\Gamma'$ merge
into a single orbit in $\Gamma''$ after reconnection in a $2-
$encounter, also agrees with (\ref{assumption}); reconnection in
an $l-$encounter with $l>2$ is reducible to a sequence of
reconnections in $2-$ encounters. By repeated activation of the
encounters resulting in joining and disjoining the orbits, we can
construct any pseudo-orbit quadruplet out of an orbit pair for
which (\ref{assumption}) is known to be correct; hence by
induction it is true for all quadruplets.

For the spin different from $1/2$, the Bolte-Harrison recurrence
relations (\ref{recur1}),(\ref{recur2}),(\ref{recur3}) differ by
the replacement of $2$ in the denominator by $2S+1$; for  integer
spins the sign in (\ref{recur1}) is plus. Otherwise the reasoning
remains unchanged with the result (\ref{assumption}) replaced by
\begin{eqnarray}\label{snot12}
C_{\Gamma\Gamma'}=\frac{    \left(-1\right)^{L-V+\nu_{\Gamma}%
-\nu_{\Gamma'}}}{(2S+1)^{L-V}},\quad \mbox{half-integer $S$;}\nonumber\\
C_{\Gamma\Gamma'}=\frac{1}{(2S+1)^{L-V}},\quad\mbox{integer $S$.}
\end{eqnarray}

\bigskip

\subsection{ Generating functions of RMT. GOE-GSE duality\label{Zirn}}

The generating function of GOE found by Zirnbauer \cite{Zirnb08}
has a
Weyl symmetric form,%
\[
Z_{\mathrm{GOE}}\left(  \hat{\varepsilon}\right)  =F_{\mathrm{GOE}}\left(
\hat{\varepsilon}\right)  +F_{\mathrm{GOE}}\left(  w\left(  \hat{\varepsilon
}\right)  \right)
\]
where $\hat{\varepsilon}=\left(  \varepsilon_{A}\varepsilon_{B}\varepsilon
_{C}\varepsilon_{D}\right)  $; $w$ interchanges $C$ and $D$. Denoting
$\overline{ad}=\varepsilon_{A}-\varepsilon_{D}$ etc and assuming
$\operatorname{Im}\overline{ab}>0$ we can write,
\[
F_{\mathrm{GOE}}\left(  \hat{\varepsilon}\right)  =e^{i\frac{1}{2}%
(\overline{ab}-\overline{cd})}\frac{\overline{ad}\,\,\,\overline{cb}}{\overline{ab}\,\,\,\overline{cd}}\left[
1+\frac{1}{2}\overline{ac}\,\,\,\overline{bd}\,\frac
{2+i\,\overline{cd}}{\overline{cd}^{2}}\,\,f_{+}\left(
\frac{\overline{ab}}{2}\right) \right]
\]
with $f_{+}$ defined above, see (\ref{deffplusminus}).

The generating function of GSE is given in \cite{Zirnb08} as,%
\begin{align*}
Z_{\mathrm{GSE}}\left(  \varepsilon\right)   & =Z_{\mathrm{GUE}}\left(
\varepsilon\right)  +\frac{\overline{ac}\,\,\,\overline{ad}\,\,\,\overline
{bc\,}\,\,\overline{bd}\,\,\,\,}{\overline{ab}\,\,\,\overline{cd}\,\,\,}%
\frac{1}{4}G_{1}\left(  \frac{\overline{ab}}{2}\right)  G_{0}\left(
\frac{\overline{cd}}{2}\right)  ,\\
G_{1}\left(  z\right)   & =i\int_{1}^{\infty}e^{izq}q\,dq=-\frac{e^{iz}}%
{z^{2}}\left(  i+z\right)  ,\\
G_{0}\left(  x\right)   & =\int_{-1}^{1}\frac{\sin qx}{q}dq=\pi\,+\int
_{1}^{\infty}\frac{e^{-iqx}}{iq}dq-\int_{1}^{\infty}\frac{e^{iqx}}{iq}dq\\
& =\left(  \pi-ie^{-ix}f_{+}^{\ast}(x)\right)  +ie^{ix}f_{+}\left(  x\right)
=G_{0}^{\left(  1\right)  }+G_{0}^{\left(  2\right)  }.
\end{align*}
Here $Z_{\mathrm{GUE}}\left(  \varepsilon\right)  $ is the generating function
of the unitary ensemble; in the last line $x=\overline{cd}/2$ is assumed real positive.

The GSE generating function can be transformed to a sum of two components
connected by the Weyl substitution $w$. First let us write $Z_{\mathrm{GSE}%
}\left(  \varepsilon\right)  =F^{\left(  1\right)  }+F^{(2)}$ with
\begin{align*}
F^{\left(
1\right)  }
&
=e^{i\frac{(\overline{ab}-\overline{cd})}{2}}\frac{\overline{ad}\,\,\,\overline{cb}}%
{\overline{ab}\,\,\,\overline{cd}}\left[
1+\frac{1\,\,}{2}\overline{ac}\,\,\overline{bd}\,\frac{\left(
2-i\,\overline{ab}\right)
}{\overline{ab}^{2}}f_{+}^{\ast}\left(
\frac{\overline{cd}}{2}\right)
\right] \\ &
+e^{i\frac{\overline{ab}}{2}}\frac{\overline{ad}\,\,\,\overline{cb}\,\,\,
\overline{ac}\,\,\overline{bd}}{2\overline{ab}\,\,\,\overline{cd}}\frac{\left(
2-i\,\overline{ab}\right)
}{\overline{ab}^{2}}i\pi
\end{align*}
and%
\[
F^{\left(  2\right)  }=e^{i\frac{(\overline{ab}+\overline{cd})}{2}}
\frac{\overline{ac}\,\,\overline{db}}{\overline{ab}\,\,\,\overline{dc}}%
\left[  1+\frac{1\,\,}{2}\overline{ad}\,\,\overline{bc}\,\frac{\left(  2-i\,\overline{ab}\right)  }{\overline{ab}^{2}%
}f_{+}\left(  \frac{\overline{cd}}{2}\right)  \right]  .
\]
The  part  $F^{\left(  1\right)  }$ proportional to $e^{i\frac
{(\overline{ab}-\overline{cd})}{2}}$ generates the non-oscillatory
component of the symplectic correlator while the part proportional
to $e^{i\frac{\overline{ab}}{2}}$ generates the correlator term
proportional to $\propto e^{i\varepsilon}$. The part $F^{\left(
2\right)  }\propto e^{i\frac {(\overline{ab}+\overline{cd})}{2}}$
is responsible for the component of the correlator $\propto
e^{i2\varepsilon}$.

Let us define the function $f_{+}\left(  -x\right)  $ for real
positive $x$, i.e., at the branch cut, as the average of the
values of the analytic function $f_{+}\left(  z\right)  $
at the lips of the cut,%
\[
f_{+}\left(  -x\right)  =\frac{f_{+}\left(  -x+i0\right)  +f_{+}\left(
-x-i0\right)  }{2},\quad x>0.
\]
Taking into account that%
\begin{align*}
f_{+}\left(  -x+i0\right)   & =f_{+}^{\ast}\left(  x\right)  ,\\
f_{+}(-x+i0)-f_{+}(-x-i0)  & =-2\pi ie^{ix},
\end{align*}
we have $G_{0}\left(  x\right)  =-if_{+}\left(  -x\right)  e^{-ix}%
+if_{+}\left(  x\right)  e^{ix}$: we have concealed $\pi$ in the
first summand. Now considering that the substitution $w$ changes
the sign of $\overline{cd}$
we obtain%
\begin{equation}
Z_{\mathrm{GSE}}\left(  \hat{\varepsilon}\right)  =F_{\mathrm{GSE}}\left(
\hat{\varepsilon}\right)  +F_{\mathrm{GSE}}\left(  w\left(  \hat{\varepsilon
}\right)  \right)  .\label{ZGSERMT}%
\end{equation}
with%
\[
F_{GSE}\left(  \hat{\varepsilon}\right)  =F^{\left(  1\right)  }%
=e^{i\frac{(\overline{ab}-\overline{cd})}{2}}\frac{\overline{ad}\,\,\,\overline{cb}}{\overline{ab}\,\,\,\overline{cd}}\left[  1+\frac{1\,\,}%
{2}\overline{ac}\,\,\overline{bd}\,\frac{2-i\,\overline{ab}}{\overline{ab}^{2}}f_{+}\left(
-\frac{\overline{cd}}{2}\right)  \right] .
\]
The duality relation now holds,%
\begin{align*}
F_{\mathrm{GSE}}\left(  \varepsilon_{A},\varepsilon_{B},\varepsilon
_{C},\varepsilon_{D}\right)   & =F_{\mathrm{GOE}}\left(  -\varepsilon
_{C},-\varepsilon_{D},-\varepsilon_{A},-\varepsilon_{B}\right)  ,\\
\operatorname{Re}\left(  \varepsilon_{A}-\varepsilon_{B}\right)   &
>0,\quad\operatorname{Im}\left(  \varepsilon_{A}-\varepsilon_{B}\right)
=+i0,\quad\\
\varepsilon_{C}-\varepsilon_{D}  & >0.
\end{align*}
\qquad\qquad\ \ \

It is instructive to compare the RMT function (\ref{ZGSERMT}) with the
semiclassical results. Taking (\ref{ZnGOE}) into account and using the
semiclassical duality, we have the following equivalence in the high-energy
limit,%
\begin{align}
F_{\mathrm{GSE}}\left(  \hat{\varepsilon}\right)   & \sim Z_{S}^{\left(
1\right)  }\left(  \hat{\varepsilon}\right)  +e^{i\left(  \varepsilon
_{A}-\varepsilon_{B}\right)  /2}\frac{\overline{ad}\,\,\,\overline
{cb}\,\,\,\overline{ac}\,\,\overline{bd}}{2\overline{ab}\,\,\,\overline{cd}%
}\frac{\left(  2i+\overline{ab}\right)  }{\overline{ab}^{2}}\pi
,\label{SagainsGSE}\\
F_{\mathrm{GSE}}\left(  w\left(  \hat{\varepsilon}\right)  \right)   & \sim
Z_{S}^{\left(  2\right)  }\left(  \hat{\varepsilon}\right)  \equiv
Z_{S}^{\left(  1\right)  }\left(  w(\hat{\varepsilon})\right)  .\nonumber
\end{align}
We stress that the asymptotics of $F_{\mathrm{GSE}}\left(  w\left(
\hat{\varepsilon}\right)  \right)  $ is not simply the Weyl-transposed
asymptotics of $F_{\mathrm{GSE}}\left(  \hat{\varepsilon}\right)  $. The
reason is the Stokes phenomenon; the Weyl operation changes the sign of
$\overline{cd}$ , and whereas the asymptotics of $f_{+}\left(  \overline
{cd}/2\right)  $ is purely power-like, $f_{+}\left(  -\overline{cd}/2\right)
$ contains an additional exponential summand, in accordance with%

\begin{align*}
f_{+}\left(  x\right)   & \sim\sigma_{+}\left(  x\right)  ,\quad f_{+}\left(
-x\right)  \sim i\pi e^{ix}+\sigma_{+}\left(  -x\right)  ,\\
\quad x  & \rightarrow+\infty.
\end{align*}
where $\sigma_{+}$ is the asymptotic power series  defined in
(\ref{sigmadef}).

\bigskip
\bibliographystyle{unsrt}

\bibliography{pbraun4}
\bigskip

\bigskip

\bigskip
\end{document}